\documentclass[preprint2]{aastex}
\begin{document}
\title{Investigating the Sloan Digital Sky Survey Cataclysmic Variable
SSDS J132723.39+652854.2\footnotemark} 

\author{Michael A. Wolfe\altaffilmark{2}, Paula Szkody\altaffilmark{2},
Oliver J. Fraser\altaffilmark{2}, Lee Homer\altaffilmark{2}, 
Sam Skinner\altaffilmark{2}, 
Nicole M. Silvestri\altaffilmark{2}}


\begin{abstract}

Photometric and spectroscopic observations reveal the Sloan Digital Sky
Survey cataclysmic variable SDSS J132723.39+652854.2 to be
a likely SW Sex star with an orbital period of P = 3.28 hours. 
The Sloan Digital Sky Survey spectrum shows strong He\,I, 
He\,II and Balmer emission lines. Time resolved
spectra acquired at Apache Point Observatory reveal absorption-line
structure within the emission lines near phase 0.5. 
Photometric data obtained at 
Manastash Ridge Observatory using Harris $V$ and $B$ filters reveal a high 
inclination system (i = 80$^{\circ}$) with deep ($\sim$ 2 mag.) eclipses.  
The orbital period and the spectral variations favor an 
SW Sex interpretation.
\end{abstract}

\keywords{stars: individual (SSDS J132723.39+652854.2) --- stars: novae, 
cataclysmic variables --- techniques: photometric --- techniques: 
spectroscopic}

\section{Introduction}

Cataclysmic variables (CVs) are binary star systems that consist of a 
white dwarf
(primary) and a late spectral type M star (secondary) that fills its Roche
Lobe, allowing mass to transfer to the white dwarf via the inner Lagrangian 
Point. The actual scenario of mass transfer depends on the 
magnetic field strength of the primary (Warner 1995). If the magnetic field 
is weak ($<$ 1 MG), then an accretion disc forms, whereas if the 
magnetic field is $\sim$ 1-10 MG the inner parts of the accretion disc are 
truncated where the magnetic pressure dominates.
\addtocounter{footnote}{0}\footnotetext
{Based on observations with the Apache Point
Observatory 3.5 m telescope and the Sloan Digital Sky Survey, which are owned 
and operated by the 
Astrophysical Research Consortium (ARC), and the Manastash Ridge Observatory
0.76 m telescope owned and operated by the University of Washington.
Observations were also done with the Steward Observatory 2.2 m telescope.}
\addtocounter{footnote}{1}\footnotetext{Department of Astronomy, University of Washington,
Box 351580, Seattle, WA 98195}
When the magnetic field strength exceeds $\sim$ 20 MG,
the accretion disc is totally disrupted
and mass flows along the magnetic field lines. As the above 
descriptions imply, cataclysmic variables come in an assortment of ``flavors" 
as described by Hellier (2001) and Warner (1995), with the highest field
systems called Polars and the intermediate field systems called Intermediate
Polars (IPs). SW Sex stars are one subset
of cataclysmic variables that are usually found as high 
inclination systems with orbital periods of 3-4 hrs and approximately 20 such 
systems are known\footnote{See list of SW Sex stars by Hoard at http://spider.ipac.caltech.edu/staff/hoard/research/swsex
/biglist.html}.
\begin{figure*}[!htb]
\resizebox{.8\textwidth}{!}{\plotone{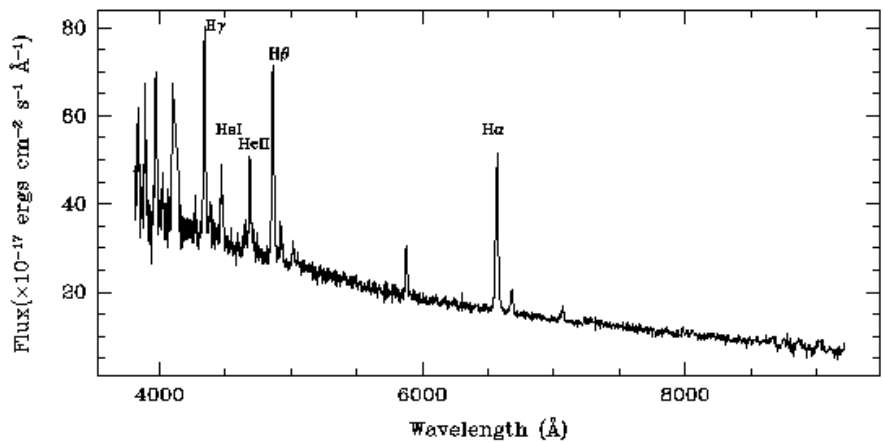}}
\caption {SDSS spectrum of SDSS J1327, a 45 min
exposure obtained 2001 Mar 20. The spectral resolution is R $\sim$ 1800.}
\end{figure*}
There are 58 Novalikes (NLs) in this period regime of which 34\% are known to be 
SW Sex stars and 17\% are known to be IPs (Ritter \& Kolb 2003). 
The SW Sex stars show an offset of $\sim$ 0.20
between photometric and spectroscopic phases, strong single peaked 
He\,I and Balmer emission lines, unusually strong He\,II (near the strength 
of H$\beta$) and absorption near $\phi$ $\sim$ 0.50 in the Balmer and He\,I 
lines (Honeycutt, Schlegel \& Kaitchuck 1986; Szkody 
\& Pich\'{e} 1990; Thorstensen et al. 1991; Hoard \& Szkody 1997). Although most 
of the 9 SW Sex stars that have been observed show low circular polarization (from
measurements with an accuracy of a few tenths of a percent; Stockman et al. 1992),
a few do have observed marginal polarization, (Rodr\'{\i}guez-Gil et al. 2001, 2002)
indicating that magnetic fields may play a role in these systems and they may 
indeed all be IPs.
\section{Follow-up Observations}
\begin{deluxetable}{lccccc}
\tablenum{1}
\tablewidth{0pt}
\tablecaption{Summary of Follow-up Observations}
\tablehead{
\colhead{UT Date} & \colhead{UT} & \colhead{Obs} & \colhead{Data} &
\colhead{Exp(s)} & \colhead{Num}}
\startdata
2001 Mar 20 & 9:35 & SDSS & Spectrum & 2701 & 1 \\
2002 May 11 & 4:25 - 5:32 & SO & Polarimetry & 800 & 5 \\
2002 Aug 01 & 5:02 - 9:23 & MRO & $V$ Photometry & 360 & 41 \\
2002 Aug 09 & 5:26 - 9:06 & MRO & $V$ Photometry & 210 & 53 \\
2002 Aug 10 & 4:56 - 9:19 & MRO & $B$ Photometry & 360 & 42 \\
2003 Jan 10 & 9:14 - 13:19 & APO & Spectra & 300 & 43 \\
\enddata
\end{deluxetable}

Recently, several new CVs have been discovered in the Sloan Digital Sky Survey
(SDSS) and are available in the Early Data Release 
(Szkody et al. 2002; Paper 1). Paper 1 describes the  
SDSS (Fukugita et al. 1996; Gunn et al. 1998; York et al. 2000; Hogg et al. 2001;
Smith et al. 2002; Stoughton et al. 2002; Pier et al. 2003), 
the methods for discerning cataclysmic variables from 
other celestial objects, and identifies 19 new CVs from the spectra 
obtained through December 2000. Light curves and time resolved spectra are 
being acquired for these objects to identify their orbital periods and 
determine their type of CV. Paper 2 (Szkody et al. 2003) identifies an 
additional 42 CVs from SDSS data through December 2001. The CV 
SDSS J132723.39+652854.2 (hereafter SDSS J1327) is one object from Paper 2.
The SDSS spectrum of SDSS J1327 (Figure 1) shows the typical strong single peaked 
He\,II $\lambda$ 4686 emission line that is indicative of an SW Sex star. 
This paper reports the results of time-series photometry, spectroscopy, 
spectropolarimetry and Doppler tomography that support the identification of
this system as an SW Sex star.
\begin{figure}
\plotone{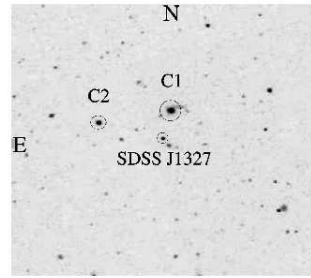}
\caption {Finding chart ($5\arcmin\times5\arcmin$) for SDSS J1327 obtained from
the Sloan Digital Sky Survey. The coordinates are $\alpha$(J2000) = 
13$^{\rm h}$27$^{\rm m}$23.39$^{\rm s}$ and $\delta$(J2000) = 
65$^{\circ}28\arcmin54.29\arcsec$. The two 
comparison stars are labeled C1 and C2 and the object is labeled SDSS J1327. The 
Sloan filter used is $g$.}
\end{figure}
Photometric observations using Manastash Ridge Observatory's
(MRO) 0.76 m telescope obtained data in the Harris $V$ and $B$ filters
with a 1024 $\times$ 1024 Ford Aerospace CCD in a Joule-Thompson dewar cooled
to $-$112$^{\circ}$ C by a Cryotiger cooling system. Photometry was acquired on 
three different nights in 2002 Aug with a time-resolution of 360 s and 210 s 
in the $V$ filter and 360 s in the $B$ filter (see Table 1).
Data were reduced using
IRAF \footnote{{IRAF (Image Reduction and Analysis Facility) is distributed 
by the National Optical Astronomy Observatory, which is operated by 
Associations of Universities for Research in Astronomy, Inc., under agreement 
with the National Science Foundation.}} routines and ``qphot" determined the 
magnitudes of the comparison stars (Figure 2) and SDSS J1327. The magnitudes were 
then used to produce differential light 
curves in each filter. The out-of-eclipse magnitude of SDSS J1327 in the 
SDSS green filter is $g$ = 17.77 $\pm$ 0.01.
Figure 3 shows the differential $B$ and $V$ light curves of 
SDSS J1327 with respect to C1. The differential light curves of the 2 comparison
stars are constant to 0.05 mag in $B$ and 0.08 mag in $V$ during the same 
intervals.

Spectroscopic data were taken with the 3.5 m telescope at Apache 
Point Observatory (APO). Forty-three time-resolved (300 s) spectra were obtained
using the Double Imaging Spectrograph with a 1.5$\arcsec$ slit   
in a high resolution mode with 2$\times$1 binning resulting in a 
resolution of 1.9 \AA\ in the blue (4000-5250 \AA) and 
2.5 \AA\
in the red (5800-7000 \AA). 
These spectra were reduced with IRAF spectroscopic reduction packages. The 
spectra were combined by twos to reduce the noise and these combined spectra 
were used in producing the radial velocity curves. However, the time-resolution
of the 43 individual spectra were utilized to calculate the flux, continuum 
and equivalent width. The software package ``e" in splot was used to calculate 
the values for the flux, continuum and equivalent width. 

Spectropolarimetry was obtained at Steward Observatory on 2002 May 11
using the polarimeter on the 2.2 m telescope with a resolution of 15 \AA\ 
between 4000 \AA\ and 8000 \AA. Five observations giving a total 4000 s exposure 
resulted in a polarization 
limit of 0.24\% $\pm$ 0.17\%. See Table 1 for a summary of the above observations.

\section{Discussion}
\subsection{Photometry}

Deep eclipses of $\sim$ 2.0 magnitudes in the $B$ filter and $\sim$ 1.8 
magnitudes 
in the $V$ filter are apparent in the light curves of SDSS J1327 (Figure 3).
\begin{figure*}[!htb]
\resizebox{.8\textwidth}{!}{\plotone{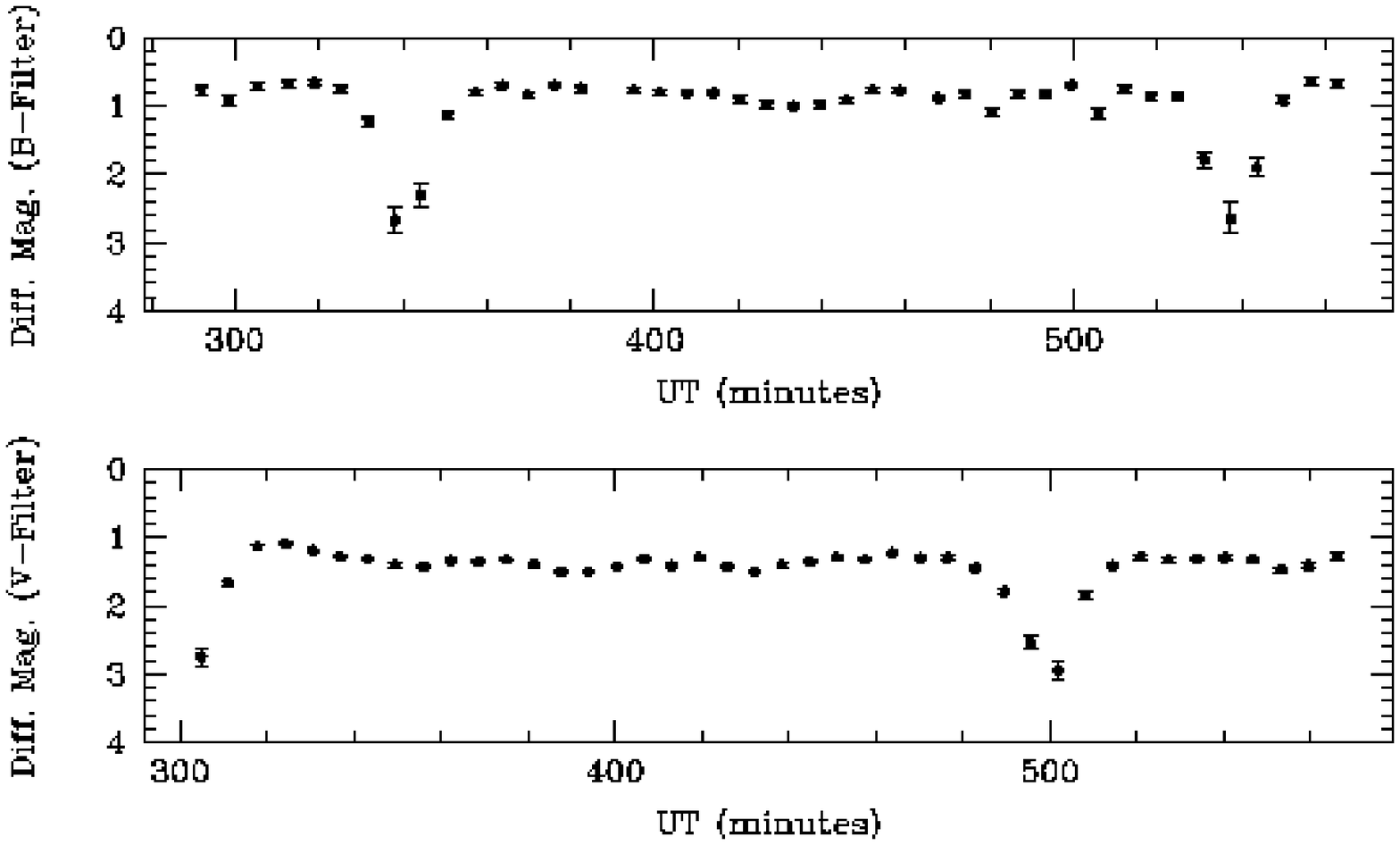}}
\caption {MRO light curves of SDSS J1327 in $B$ filter 2002 Aug 10 (top) and $V$ 
filter 2002 Aug 1 (bottom).}
\end{figure*}
These depths and the change with color indicate that the white dwarf and
the inner disc is
eclipsed in this system. The light curves show no evidence of a pre-eclipse
hump that would signify a strong hot spot.

The timings of the eclipses are listed in Table 2.
\begin{deluxetable}{lcc}
\tablenum{2}
\tablewidth{0pt}
\tablecaption{Eclipse Timings}
\tablehead{
\colhead{UT} & \colhead{UT} & \colhead{HJD 2452+}\\
\colhead{(Date)} & \colhead{(Minutes)} & \colhead{}}
\startdata
2002 Aug 01 & 305$\pm$3 & 487.71639$\pm$0.002 \\
2002 Aug 01 & 496$\pm$3 & 487.84203$\pm$0.002 \\
2002 Aug 09 & 402$\pm$2 & 495.77664$\pm$0.001 \\
2002 Aug 10 & 335$\pm$3 & 496.73010$\pm$0.002 \\
2002 Aug 10 & 533$\pm$3 & 496.86760$\pm$0.002 \\
2003 Jan 10 & 580$\pm$3 & 649.90442$\pm$0.002 \\
2003 Jan 10 & 778$\pm$3 & 650.04192$\pm$0.002 \\
\enddata
\end{deluxetable}
The coverage between Aug 1 and Aug 10 provides an unambiguous cycle count
during this interval and results in the eclipse ephemeris:
\begin{eqnarray}
\rm T_{0} = 2,452,487.7164 HJD \pm 0.0036 \nonumber \\ + 0.136647E \pm 0.000062
\end{eqnarray}
This ephemeris was used to calculate all phases for this paper.

The period of 3.28 hrs places SDSS J1327 in the regime of SW Sex stars and
IPs (Warner 1995). The mass transfer rates are usually large in this period
range, resulting in prominent accretion discs and low hot spot contribution
to the total light. 

Since the time-resolution of the photometry is only six minutes, it was not
possible to detect any structure within the eclipse nor search for any
short timescale variations that could be related to a spin period for the
white dwarf. 

\subsection{Spectroscopy}
\begin{figure*}[!htb]
\resizebox{.8\textwidth}{!}{\plotone{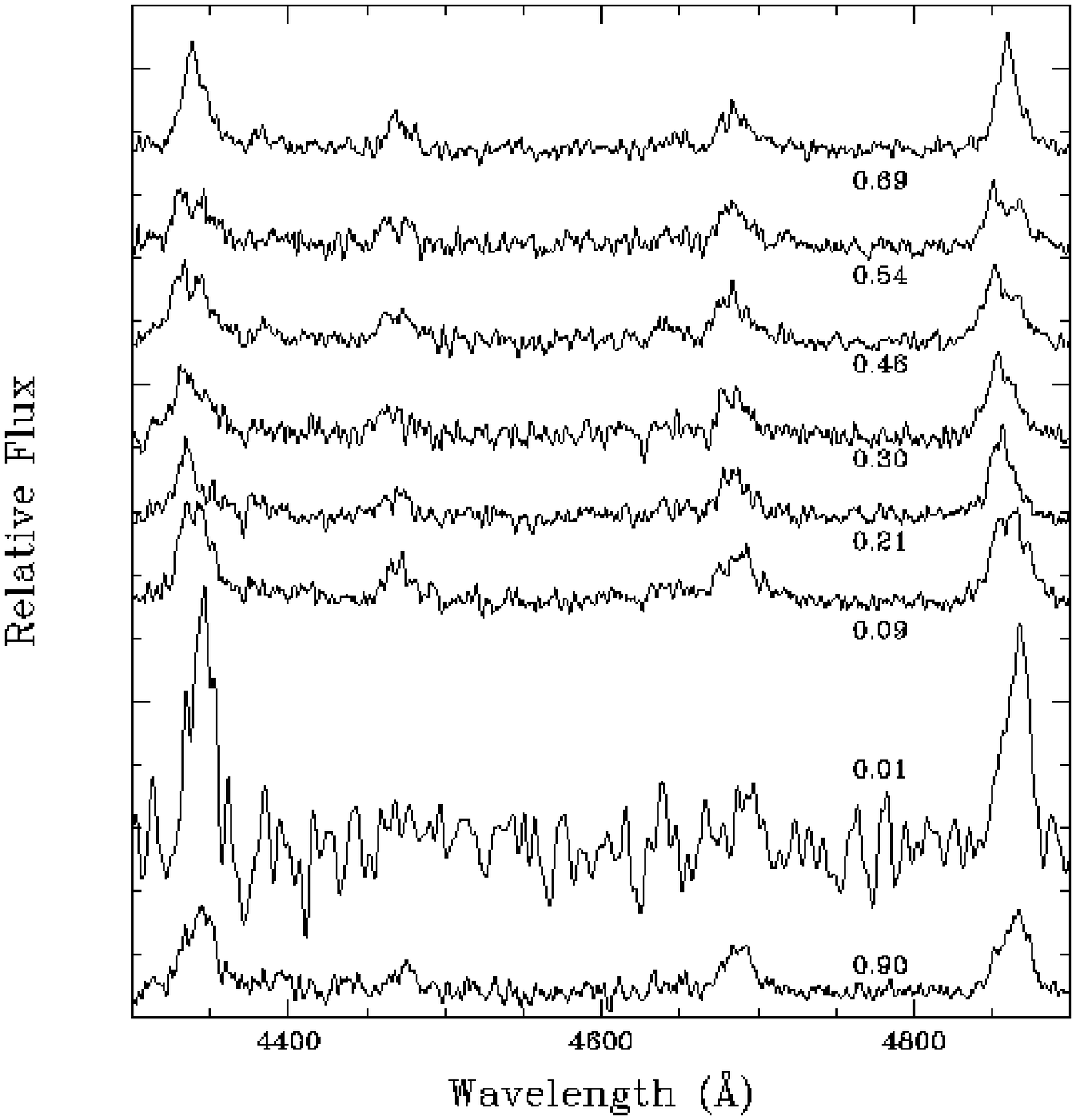}}
\caption {Sample of APO spectra as a function of
phase showing emission line structure of H$\beta$, He\,II, He\,I and H$\gamma$.
All spectra have been normalized to a continuum level of 1.}
\end{figure*}
The 43 time-resolved spectra acquired at APO reveal different eclipse
depths in the lines and a changing structure within the lines during the
orbit. Figure 4
shows a sample of the normalized spectra. This Figure, as well
as the equivalent width and flux plots (Figure 5) reveal that the
He\,II flux has a deeper eclipse than H$\beta$.
This indicates an origin for
the He\,II that is close to the white dwarf and hence, eclipsed, while the
Balmer lines remain uneclipsed. 
\begin{figure*}[!htb]
\resizebox{.8\textwidth}{!}{\plotone{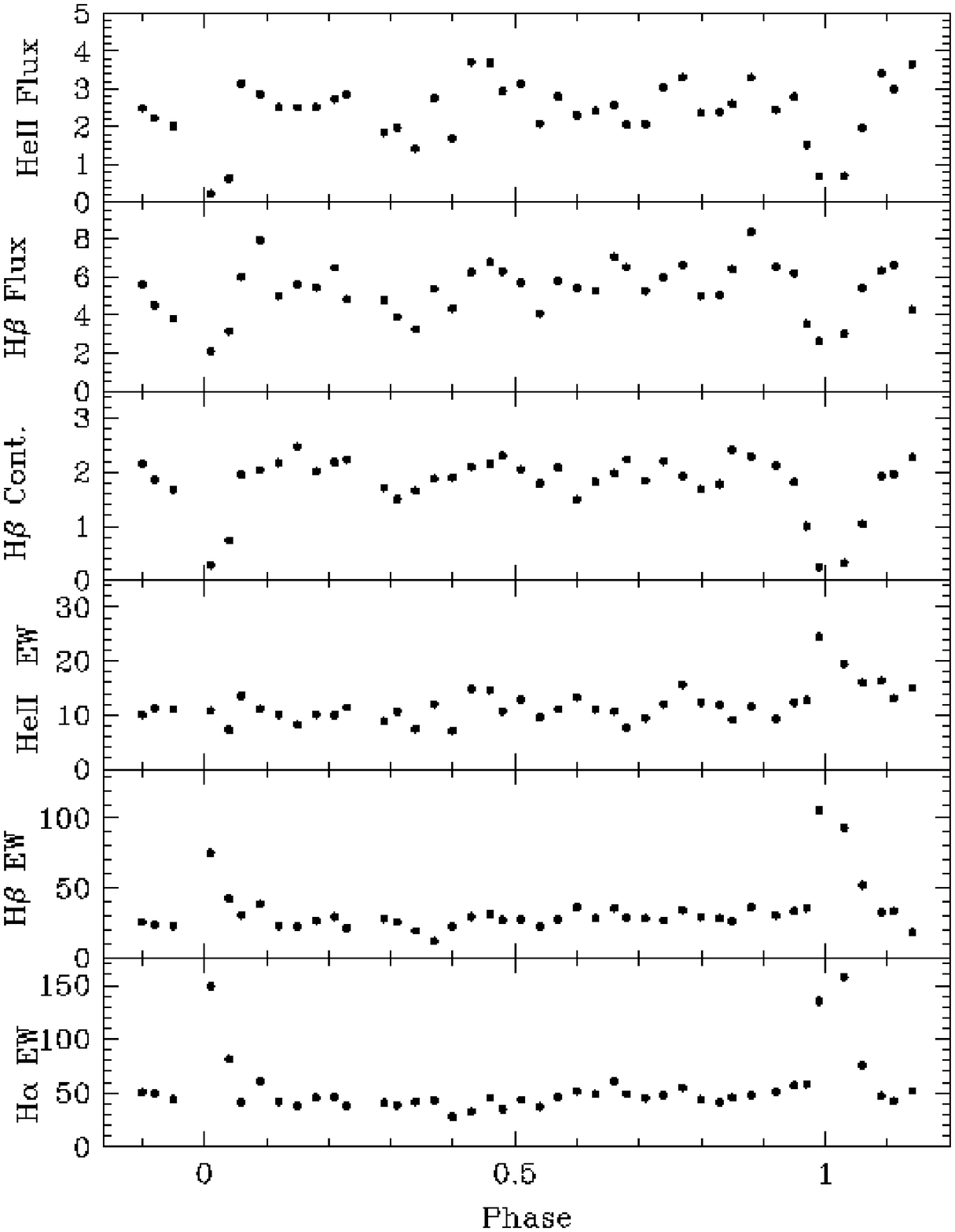}}
\caption {Flux and equivalent width (EW) plots for H$\beta$, He\,II and 
H$\alpha$.
Units of flux for H$\beta$ and He\,II are 10$^{-15}$ ergs 
cm$^{-2}$ s$^{-1}$. The units of the H$\beta$ continuum are 
10$^{-16}$ ergs cm$^{-2}$ s$^{-1}$
\AA$^{-1}$. Units of EW are \AA. The error in all plots is approximately 10\% of 
the values of the points.}
\end{figure*}
This is typical for disc systems at high
inclination (Downes et al. 1986). The time-resolved spectra also show
absorption cores in the Balmer and He\,I lines. This absorption is
apparent from phases 0.43-0.57 and is strongest at phase 0.54.
This absorption is an indication 
of a SW Sex star (Szkody \& Pich\'{e} 1990), however, 
it is not as deep as in other SW Sex stars such as DW UMa 
(Szkody \& Pich\'{e} 1990), BH Lyn (Hoard \& Szkody 1997) and V1315 Aql
(Dhillon, Marsh \& Jones 1991). In these stars absorption can be  
below the continuum in the higher order Balmer and He\,I lines.

\begin{deluxetable}{lcccccc}
\tablenum{3}
\tablewidth{0pt}
\tablecaption{Radial Velocity Solutions}
\tablehead{
\colhead{Line} & \colhead{$\gamma$} & \colhead{K} &
\colhead{$\phi$ offset \tablenotemark{a}} & \colhead{$\sigma$ \tablenotemark{b}}\\
\colhead {} & \colhead{(km s$^{-1}$)} & \colhead{(km s$^{-1}$)} & \colhead{} & 
\colhead{(km s$^{-1}$)}}
\startdata
H$\alpha$ & $-$41$\pm$3 & 299$\pm$16 & 0.20$\pm$0.01 & 43 \\
H$\beta$ & $-$69$\pm$2 & 188$\pm$18 & 0.18$\pm$0.02 & 55 \\
He\,II & $-$69$\pm$3 & 163$\pm$18 & 0.22$\pm$0.02 & 46 \\
\tablenotetext{a}{This is the offset between photometric and spectroscopic phases.}
\tablenotetext{b}{$\sigma$ is the overall error of the best sine fit. See text for 
the sine function.}
\enddata
\end{deluxetable}
Radial velocities were measured using a double Gaussian methodology (Shafter
1985). FWHM values of the individual Gaussians of the 60-80 km s$^{-1}$ 
and the Gaussian separations of 1400-1800 km s$^{-1}$ resulted in 
velocities giving the best fit (least overall error $\sigma$) to the sine function
\begin{equation}
\rm v = \gamma - K sin(2\pi(\phi - \phi_0))
\end{equation}
 where $\gamma$ is the systemic
velocity, K is the semi-amplitude and $\phi$ is the offset from the
photometric eclipse phasing (inferior conjunction of the secondary).
The best fit parameters for
the lines of H$\alpha$, H$\beta$ and He\,II are also listed in Table 3 and
the radial velocity curves are shown in Figure 6.
The phase offset between the 
photometry and spectroscopy is large (0.2) and provides further evidence
that SDSS J1327 is an SW Sex star. 
\begin{figure*}[!htb]
\resizebox{.8\textwidth}{!}{\plotone{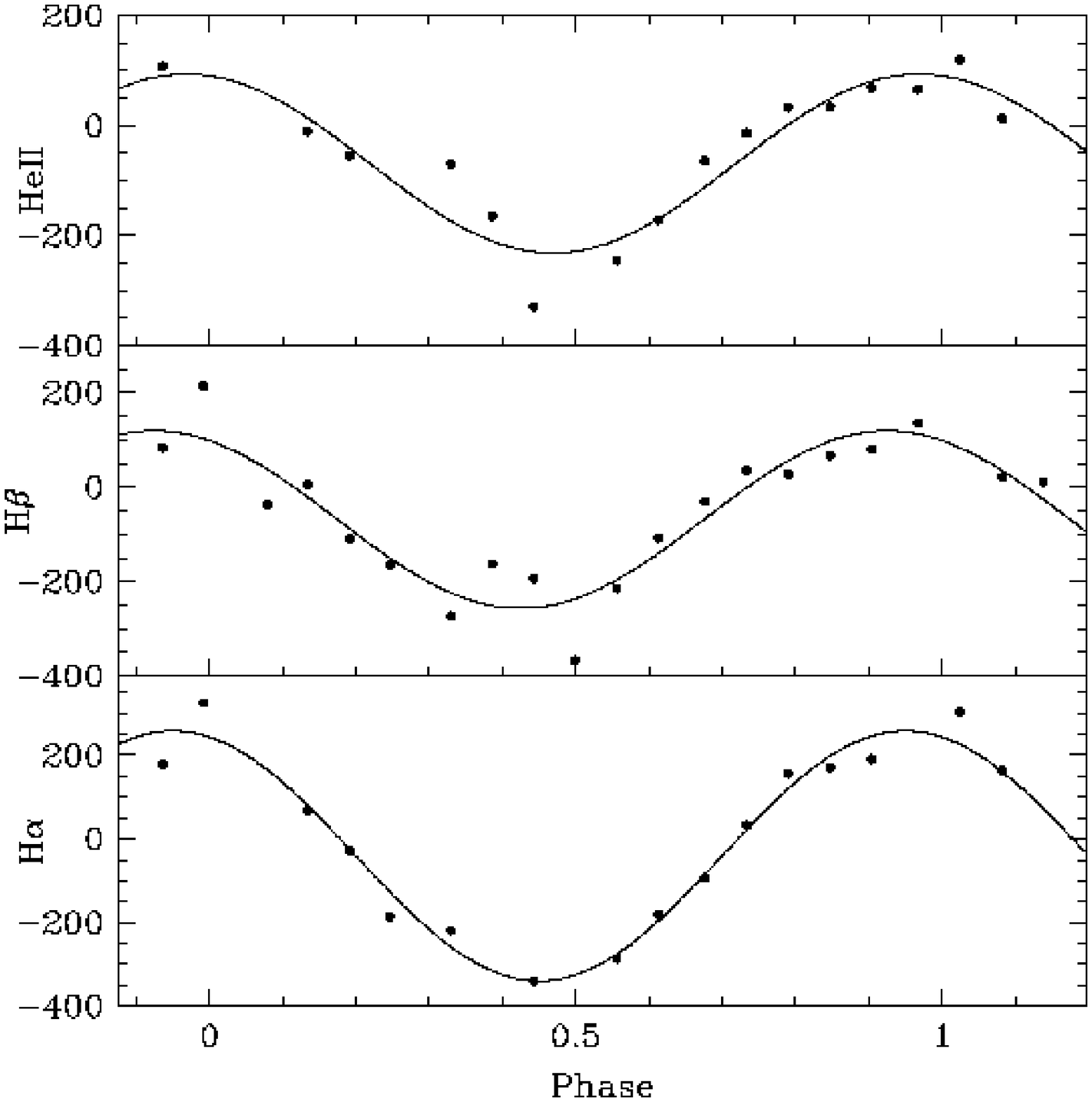}}
\caption {Radial velocity curves for SDSS J132723.39+652854.2 in km s$^{-1}$.}
\end{figure*}

System parameters of M$_{\rm s}$, M$_{\rm wd}$, q and i can be estimated from 
the combination of the photometric and spectroscopic data, 
using equations (4), (5) and (7) from Downes et al.
(1986) and using $\beta$ = 1 (no 
deviation from the standard spherical approximation to the Roche geometry at
small q).   
Using the value of $\phi_{1/2}$ = 0.045 (the orbital phase at which
50\% of the eclipsed star's light is reduced), K$_{\rm wd}$ = 163 km s$^{-1}$
(from He\,II) and the determined period, P = 3.28 hrs, yields:
\begin{center}
i = 80$^{\circ}$ +3$^{\circ}$/$-$1$^{\circ}$,\\
M$_{\rm s}$ = 0.4 $\pm$ 0.2 M$_{\odot}$,\\ 
M$_{\rm wd}$ = 0.8 $\pm$ 0.2 M$_{\odot}$,\\
q = 0.5 $\pm$ 0.1.
\end{center}

The radius of the secondary R$_{\rm s}$, and a (separation between the center of
mass of the primary and secondary) can also be estimated using the Patterson 
(1984)
approximation for the mass-radius relationship:
\begin{center}
R$_{\rm s}$ = 0.4 $\pm$ 0.1 R$_{\odot}$,\\ 
a = 1.3 $\pm$ 0.1 R$_{\odot}$.
\end{center}

These resulting parameters for SDSS J1327 are very similar to the values 
determined for
V1315 Aql (Downes et al. 1986).

\begin{figure*}[!htb]
\resizebox{.8\textwidth}{!}{\plotone{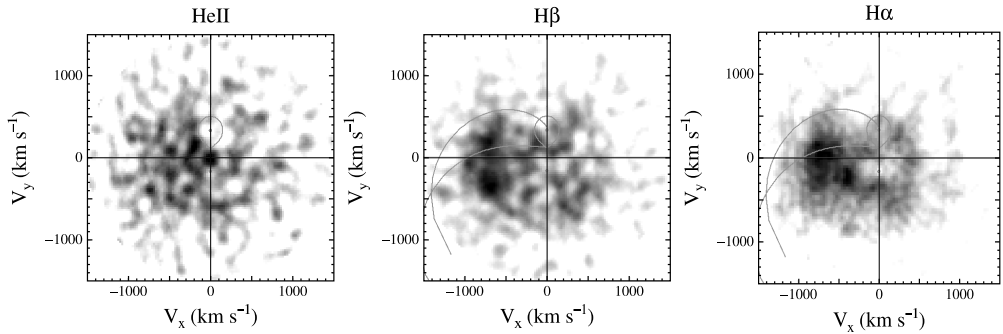}}
\caption {Doppler tomograms for SDSS 1327 from a) He\,II, b) H$\beta$ and
c) H$\alpha$. The lower curve is the trajectory of the transfer stream and the upper
curve is the Keplerian velocities in the disc along the stream trajectory.}
\end{figure*}
Doppler tomograms for H$\alpha$, H$\beta$ and He\,II were also constructed 
in order to further locate the origin of the emission lines. 
The Fourier-filtered back-projection routines of Horne (1991) were used to
compute the tomogram, using the systemic velocity for He\,II for all the lines.
The spectra were binned into 0.04 phase intervals,
with eclipse phases 0.95-0.05 omitted.
The results are shown in Figure 7a-c, along with the
Roche lobe of the secondary and the trajectory of the transfer stream (lower
curve) and the Keplerian velocities in the disc along the stream trajectory
(upper curve) plotted for a mass ratio of 2 and the K of
He\,II (Table 2). These
tomograms show a different structure in each of the lines. The He\,II (Figure
7a) shows an emission region close to the white dwarf. 
This is consistent with the
eclipse of the He\,II lines as described above. The H$\beta$ tomogram 
(Figure 7b) shows two spots, one near the edge of the mass transfer stream
location and the other in the $-$V$_{\rm x}$, $-$V$_{\rm y}$ quadrant.
The H$\alpha$
tomogram (Figure 7c) shows a region of emission primarily on the $-$V$_{\rm x}$ 
axis,
with some evidence of a ring of emission from the disc. None
of the tomograms show any evidence for irradiation of the secondary.

These tomograms are not like those of typical disc systems (Kaitchuck et al.
1994) which show a ring of emission with a prominent hot spot in the 
$-$V$_{\rm x}$, +V$_{\rm y}$ quadrant. However, they are very similar to the 
tomograms
of other SW Sex stars (e.g.
BH Lyn, UU Aqr; Hoard \& Szkody 1997, Hoard et al. 1998).
While many models to explain the SW Sex phenomena have been invoked (see the
summary in Warner 1995), Hoard et al. (1998)
account for the odd location of the emission in tomograms
and the differences between
systems. They use a model with a stream impact that is dependent on the mass 
transfer
rate. At high rates (periods near 4 hrs), there is an explosive impact of
the stream with the disc that produces a bright spot at the impact site and
effectively stops the flow. At low mass transfer rates (periods near 3 hrs), 
the stream overflows the disc and continues downstream, creating absorption
near phase 0.5. The relatively short period of SDSS J1327 should place it
near the lower mass transfer systems such as V1315 Aql (Dhillon, Marsh \&
Jones 1991), so overflow would be expected. This could be the cause of  
the slightly displaced emission in the H$\alpha$ tomogram and  the more
prominent displaced emission 
in the H$\beta$ tomogram. 

\subsection{Polarimetry}

The spectropolarimetry measurement revealed a circular
polarization limit of 0.24\%, ruling out the presence of a strong magnetic
field in the white dwarf in SDSS J1327. However, this limit is typical for
most IPs and SW Sex stars (Stockman et al. 1992) and only 2 SW Sex stars 
have marginally significant measured 
polarization at low levels of a few tenths of a percent  
(Rodr\'{\i}guez-Gil et al. 2001, 2002). 

\section{Conclusions}

SDSS J1327 is a CV discovered in the Sloan Digital Sky Survey
which shows strong Balmer and He\,II emission lines. The light curve of SDSS J1327 
displays deep eclipses,
indicating a high inclination system with an orbital period of 3.28 hrs.
An offset of 0.2 phase between the photometric and spectroscopic phases, as
well as the presence of absorption within the He\,I and Balmer emission lines
indicates that SDSS J1327 is an SW Sex 
star. Doppler tomography confirms this picture, showing the He\,II emission
concentrated near the white dwarf and the Balmer emission concentrated in
phases downstream of the hot spot.
The low limiting value to the circular polarization of 
0.24\%  is also consistent with the available data on SW Sex stars.
While the emerging picture of this system seems clear, further photometry
at better time resolution is needed to resolve the eclipse structure and 
search for any short periodicities that could be related to the spin period
of the white dwarf.

This work gratefully acknowledges Gary Schmidt for obtaining the spectropolarimetry
measurements, Don Schneider for his thoughtful input, Scott Anderson for providing 
the SDSS finding chart for SDSS J1327 and Don Hoard for his list of SW Sex stars. 
This work was partially 
funded by the Mary Gates Endowment for Students, Washington Space
Grant NGT5-40084, NSF Grant AST-0205875 and an RRF Grant from the University
of Washington.

Funding for the creation and distribution of the SDSS Archive has been 
provided by the Alfred P. Sloan Foundation, the Participating Institutions, 
the National Aeronautics and Space Administration, the National Science 
Foundation, the U.S. Department of Energy, the Japanese Monbukagakusho, and 
the Max Planck Society. The SDSS Web site is http://www.sdss.org/. 

The SDSS is managed by the Astrophysical Research Consortium (ARC) for the 
Participating Institutions. The Participating Institutions are The University 
of Chicago, Fermilab, the Institute for Advanced Study, the Japan 
Participation Group, The Johns Hopkins University, Los Alamos National 
Laboratory, the Max-Planck-Institute for Astronomy (MPIA), the 
Max-Planck-Institute for Astrophysics (MPA), New Mexico State University, 
University of Pittsburgh, Princeton University, the United States Naval
Observatory, and the University of Washington.

\clearpage


\begin{references}

\reference{} Dhillon, V. S., Marsh, T. R., \& Jones, D. H. P. 1991, 
\mnras, 252, 342.
\reference{} Downes, R. A., Mateo, M., Szkody, P., Jenner, D.C., \& Margon, B. 
1986, \apj, 301, 240.
\reference{} Fukugita, M., Ichikawa, T., Gunn, J. E., Doi, M., Shimasaku, K.,
\& Schneider, D. P. 1996, \aj, 111, 1748.
\reference{} Gunn, J. E. et al. 1998, \aj, 116, 3040.
\reference{} Hellier, C. 2001, in Cataclysmic Variable Stars: How and Why They
Vary (Chichester: Praxis Publishing Ltd).
\reference{} Hoard D. W. \&  Szkody, P. 1997, \apj, 481, 433.
\reference{} Hoard, D. W., Still, M. D., Szkody, P., Smith, R. C., \&
Buckley, D. A. H. 1998, \mnras, 294, 689.
\reference{} Hogg, D. W., Finkbeiner, D. P., Schlegel, D. J., \& Gunn, J. E.
2001, \aj, 122, 2129.
\reference{} Honeycutt, R. K., Schlegel, E. M., \& Kaitchuck, R. H. 1986, 
\apj, 302, 388.
\reference{} Horne, K. 1991, in Fundamental Properties of Cataclysmic 
Variables Stars, ed. A. Shafter (San Diego: Mount Laguna Obs.), 23.
\reference{} Kaitchuck, R. H., Schlegel, E. M., Honeycutt, R. K., Horne, K., 
Marsh, T. R., White II, J. C., \&  Mansperger, C. S. 1994, \apjs, 93, 519.
\reference{} Patterson, J. 1984, \apj, 54, 443.
\reference{} Pier, J. R., Munn, J. A., Hindsley, R. B., Hennessy, G. S., Kent,
S. M., Lupton, R. H., \& Ivezic, Z. 2003, \aj, 125, 1559.
\reference{} Ritter, H., \& Kolb, U. 2003, A\&A, in press (astro-ph/0301444). 
\reference{} Rodr\'{\i}guez-Gil, P., Casares, J., Mart\'{\i}nez-Pais, I. G.,
Hakala, P., \& Steeghs, D. 2001, \apj, 548, L49.
\reference{} Rodr\'{\i}guez-Gil, P., Casares, J., Mart\'{\i}nez-Pais, I. G., \&
Hakala, P. 2002, in ASP Conf. Ser. 261, The Physics of Cataclysmic Variables 
and Related Objects, eds. B. T. G\"{a}nsicke, K. Beuerman and K. Reinsch 
(San Francisco: ASP), 533. 
\reference{} Shafter, A. W. 1985, in Cataclysmic Variable and Low Mass X-Ray
Binaries, eds. D.Q. Lamb and J. Patterson (Dordrecht: Reidel), 355.
\reference{} Smith, J. A. et al. 2002, \aj, 123, 2121.
\reference{} Stockman, H. S., Schmidt, G. D., Berriman, G., Liebert, J.,
Moore, R. L., \& Wickramasinghe, D. T. 1992, \apj, 401, 628.
\reference{} Stoughton, C. et al. 2002, \aj, 123, 485.
\reference{} Szkody, P. \& Pich\'{e}, F. 1990, \apj, 361, 235.
\reference{} Szkody, P. et al. 2002, \aj, 123, 430. (Paper 1)
\reference{} Szkody, P. et al. 2003, \aj, submitted. (Paper 2)
\reference{} Thorstensen, J. R., Ringwald, F. A., Wade, R. A., Schmidt, G. D.,
\& Norsworthy, J. E. 1991, \aj, 102, 272.
\reference{} York, D. G. et al. 2000, \aj, 120, 1579.
\reference{} Warner, B. 1995, in Cataclysmic Variable Stars (Cambridge: 
Cambridge University Press).   
\end{references}
\end{document}